\documentclass[aps,amsfonts,amssymb,amsmath,preprintnumbers,superscriptaddress,twocolumn,nofootinbib,english,prl,10pt]{revtex4-1}
\usepackage[T1]{fontenc}
\usepackage[latin9]{inputenc}
\usepackage{babel}
\usepackage{amstext}
\usepackage{graphicx}
\usepackage{esint}
\usepackage{upgreek}
\usepackage[unicode=true,pdfusetitle,
 bookmarks=true,bookmarksnumbered=false,bookmarksopen=false,
 breaklinks=false,pdfborder={0 0 0},backref=false,colorlinks=false]
 {hyperref}
\begin{document}

\title{Evolution of the jet opening angle distribution in holographic plasma}

\author{Krishna Rajagopal, Andrey V. Sadofyev and Wilke van der Schee}

\affiliation{Center for Theoretical Physics, MIT, Cambridge, MA 02139, USA}

\preprint{MIT-CTP-4765}
\begin{abstract}
We use holography to analyze the evolution of an ensemble of jets, with an initial probability distribution
for their energy and opening angle as 
in
proton-proton (pp) collisions,
as 
they propagate
through an expanding cooling droplet of strongly coupled plasma as in heavy ion collisions. 
We identify two competing effects: 
(i) each individual jet widens as it propagates; 
(ii) the opening angle distribution
for jets emerging from the plasma within any specified range of energies has been 
pushed toward smaller angles, comparing to pp jets
with the same energies.
The second effect arises because
small-angle jets suffer less energy loss and because 
jets with a higher initial energy are less probable in the ensemble. 
We illustrate both effects in a simple two-parameter model, 
and find that their consequence in sum is that the opening
angle distribution for jets in any range of
energies contains fewer narrow and wide jets.
Either effect can dominate in the mean opening angle,
for not unreasonable values of the parameters. 
So,
the mean opening angle
for jets with a given energy can easily shift toward smaller angles, as experimental data
may indicate, even
while every jet in the ensemble broadens.
\end{abstract}
\maketitle

\section{Introduction}


The discovery that the plasma that filled the microseconds-old
universe and that is recreated in 
nucleus-nucleus collisions at RHIC and the LHC
is a strongly coupled liquid poses many outstanding challenges, including understanding
how it emerges from an asymptotically free gauge theory that is weakly coupled at short distances.  
This longer term goal requires understanding how probes of
the plasma produced in hard processes in the same collision interact with the plasma,
so that measurements of such probes can be used to discern the structure
of the plasma as a function of resolution scale.  Energetic jets are particularly
interesting probes because their formation and subsequent evolution 
within the plasma involve
physics at many length scales.

Although a holographic plasma is strongly coupled at all length scales
rather than being asymptotically free, because calculations done via their dual
gravitational description can be used to gain reliable understanding of highly
dynamical processes at strong coupling these theories have been used
to provide benchmarks for 
various aspects of the dynamics of hard probes
propagating through strongly coupled plasma~\cite{Herzog:2006gh,Liu:2006ug,CasalderreySolana:2006rq,Gubser:2006bz,Liu:2006nn,Chernicoff:2006hi,Gubser:2006nz,CasalderreySolana:2007qw,Chesler:2008uy,Gubser:2008as,Arnold:2010ir,Arnold:2011qi,Chesler:2011nc,Ficnar:2013wba,Chesler:2013cqa,Ficnar:2013qxa,Chesler:2014jva,Morad:2014xla,Chesler:2015nqz,Casalderrey-Solana:2015tas}.  
We shall focus on the proxies for light quark jets analyzed in Refs.~\cite{Chesler:2008uy,Chesler:2008wd,Chesler:2014jva,Chesler:2015nqz},
introducing them 
into hydrodynamic droplets of plasma whose expansion and cooling resembles that
in  heavy ion collisions with zero impact parameter, rather than static (slabs of) plasma with a constant temperature.  For the first time,
we shall analyze an ensemble of such jets with a distribution of jet energies and jet opening angles taken
from a perturbative QCD description of jet production in pp collisions.  
We analyze how this perturbative QCD distribution 
is modified via tracking how an ensemble of jets in a holographic theory (${\cal N}=4$ supersymmetric
Yang-Mills theory) evolves as the jets propagate through an expanding and cooling droplet of 
strongly coupled plasma in that theory, in so doing gaining qualitative insights into how this distribution
may be modified in heavy ion collisions, where jets propagate through quark-gluon plasma.
(See Refs.~\cite{Casalderrey-Solana:2014bpa,Casalderrey-Solana:2015vaa} for a quite different
way to combine weakly coupled calculations of jet production and fragmentation with
a holographic, strongly coupled, calculation of parton energy loss into a hybrid model
for jet quenching.)


We know from Ref.~\cite{Chesler:2015nqz} how the energy and opening
angle of an individual jet evolves as it propagates in the strongly coupled
${\cal N}=4$ SYM plasma, at constant temperature.  A striking result
from this calculation is that all jets with the same initial opening angle (i.e.~which 
would have had the same opening angle if they had been produced in vacuum
instead of in plasma) that
follow the same trajectory through the plasma suffer the same fractional
energy loss, regardless of their initial energy.  This highlights the role that
the opening angle of a jet plays in controlling its energy loss, a qualitative
feature also seen very recently
in a weakly coupled analysis of jet quenching in QCD~\cite{Milhano:2015mng},
where it can be understood by noting that
jets with a larger initial opening angle are jets that have fragmented into more partons, 
and in particular into more resolved subjet structures, each of which loses energy as it passes through the plasma~\cite{CasalderreySolana:2012ef}.
The strong dependence of jet energy loss on jet opening angle seen
in these analyses
shows that the modification of the jet energy distribution due to propagation through the plasma
cannot be analyzed in isolation: we must analyze an ensemble of jets with
a distribution of both energy and opening angle. 


\begin{figure*}[t]
\begin{centering}
\includegraphics[width=18cm]{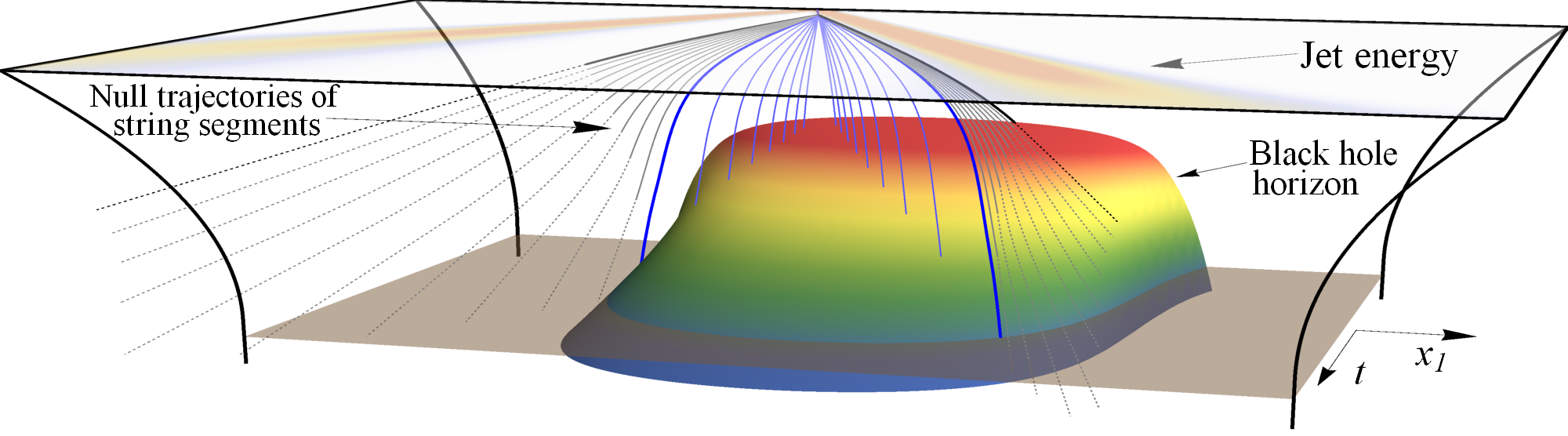}
\par\end{centering}
\vspace{-0.2cm}
\caption{An 
event where  two jets are produced at 
$x_{1}=-3.0$ fm, moving in the $\pm x_{1}$ directions, with
the same initial energy $E^{\rm init}_{\text{jet}}=100$~GeV
and with the string endpoints (heavier grey curves) moving downward into the AdS bulk 
with initial 
angles $\sigma_{0}=0.025\,(0.01)$ for the left (right) moving jet.
The colored profile is the black hole horizon in the AdS bulk, whereby both the height of the surface and 
its color indicate the temperature as the droplet of plasma expands and cools over time.  The droplet  is circularly symmetric
in the $(x_1,x_2)$ plane; $x_2$ is not shown. 
Bits of string follow  the grey and blue null trajectories.
The lower plane corresponds to the
freeze-out temperature; after the temperature drops below this value, we propagate the jet 
in vacuum and show the constant-$\sigma$ null rays
as dashed. 
The blue null rays are those that fall into the horizon 
before freeze-out; the energy
propagating along these trajectories is lost from the jet.  
The heavier blue curve shows the last packet of energy to fall into the horizon, at freeze-out.
The opening angle of the jets increase as they
traverse the plasma (e.g. for the right-moving jet
$\sigma_{*}=0.044$, almost 5 times
wider than its initial angle) as can be seen from the energy
density depicted at the boundary.\label{fig:model}}
\vspace{-0.2cm}
\end{figure*}

We discern two competing effects.  First,
as shown for constant-temperature plasma in Refs.~\cite{Chesler:2014jva,Chesler:2015nqz},
the opening angle of every individual jet in the ensemble widens as it propagates
through the plasma.  
The second 
effect arises because
the initial distribution of energies is a rapidly falling function of energy.  This means that after
the jets have propagated through the plasma, it is more likely that jets
with a given final energy are those that started with only a little more energy and lost
little energy rather
than being those which started with a much higher energy and lost a lot. 
Since the narrowest jets  lose the least energy~\cite{Chesler:2015nqz}, 
propagation 
through the plasma should push 
the opening angle distribution of jets with a given energy toward smaller
angles.  
Jets that start out with larger opening angles
get kicked down in energy, and become numerically insignificant in the ensemble.



\section{The Model}

The study of jets in a holographic plasma amounts to the evolution
of strings in an anti-de-Sitter (AdS) black hole spacetime with one extra dimension.
A pair of light quarks is represented holographically by an open fundamental
string in AdS \cite{Karch:2002sh} that is governed by
the Nambu-Goto action
$S=-T_{0}\int d\tau d\sigma\sqrt{-h}$,
with $T_{0}=\sqrt{\lambda}/2\pi$ the string tension, with $\lambda$
the 't~Hooft coupling, $\tau,$ $\sigma$ the string worldsheet coordinates
and $h_{ab}=g_{\mu\nu}\partial_{a}X^{\mu}\partial_{b}X^{\nu}$ the
string worldsheet metric, which depends on the metric of the bulk
AdS, $g_{\mu\nu}$. 

We shall follow Refs.~\cite{Chesler:2014jva,Chesler:2015nqz}
and choose strings that originate at a point at the boundary of AdS, initially propagate as if they 
were in vacuum~\cite{Chesler:2008wd}, and have sufficient energy that they can propagate
through the plasma over a distance 
$\ggg 1/T$.  As discussed in Ref.~\cite{Chesler:2015nqz},
after a time 
${\cal O}(1/T)$ initial transient
effects have fallen away. (Literally, in the gravitational description: they fall into the horizon.
In the gauge theory, gluon fields around the jet creation event are excited and we need
to wait for the jet to separate from gluon fields that are not part of the jet.)
After this time, the string 
has reached
a steady-state regime in which its 
worldsheet is approximately null and its 
configuration is specified by two parameters, corresponding in the boundary
theory to the initial energy and opening angle of the jet.  The endpoint of the string
follows a trajectory that initially angles down into the gravitational bulk with an
angle $\sigma_0$, see Fig.~\ref{fig:model}.  The initial opening angle of the jet
in the boundary gauge theory is (up to few percent
corrections) proportional to $\sigma_0$~\cite{Chesler:2015nqz}.
Once the string is in the steady-state regime, 
the energy density along the bit of
the string with initial downward angle into the bulk $\sigma$ is given by~\cite{Chesler:2014jva,Chesler:2015nqz}
\begin{equation}
e(\sigma)=\frac{A}{\sigma^{2}\sqrt{\sigma-\sigma_{0}}},\label{eq:energyprofile-1}
\end{equation}
where the constant $A$ specifies the initial energy of the jet when it enters the steady-state regime,
with $E_{\rm jet}^{\rm init} \propto A \sigma_0^{-3/2}$
for $\sigma_0\ll 1$.
($A$ is related to the $E_0$ of Ref.~\cite{Chesler:2015nqz} by $E_0=32 \,\pi^{11/2}A/\Gamma(\frac{1}{4})^6$.)
Ideally, we should initialize our strings at a point at the boundary of AdS at $t=0$ and the initial phase of the calculation should encompass
a collision, hydrodynamization of the bulk matter produced therein and, simultaneously, the initial transient dynamics of
the string. 
Details of these early dynamics 
are not relevant to the qualitative points we wish to make. For simplicity,
we shall initialize our strings at a point at the boundary of AdS at $t=1$~fm$/c$ (when
the bulk matter has hydrodynamized), use the steady-state configuration (\ref{eq:energyprofile-1})
to model the energy density on the string 
at $t=1$~fm$/c$
for all $\sigma$ from $\sigma_0$ to $\pi/2$, and take  
$E^{\rm init}_{\rm jet}\equiv \int_{\sigma_{0}}^{\pi/2}d\sigma\,e$ as our simplified definition.
To specify an ensemble of jets with some distribution of
initial energies and opening angles, we must specify an ensemble of strings 
with the appropriate distribution of $A$ and $\sigma_0$.


\begin{figure*}[t]
\begin{centering}
\includegraphics[width=6.3cm]{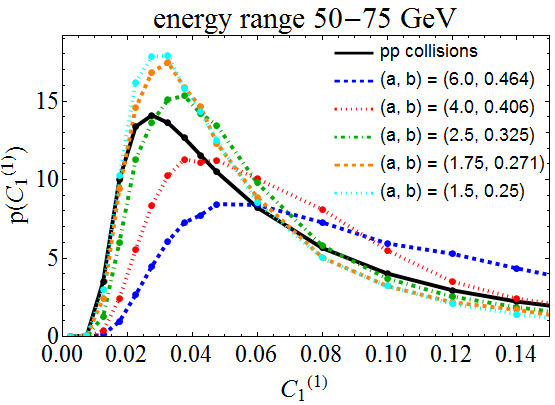}\includegraphics[width=5.9cm]{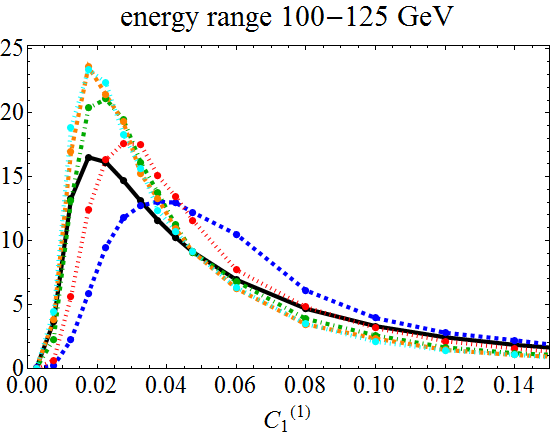}\includegraphics[width=5.9cm]{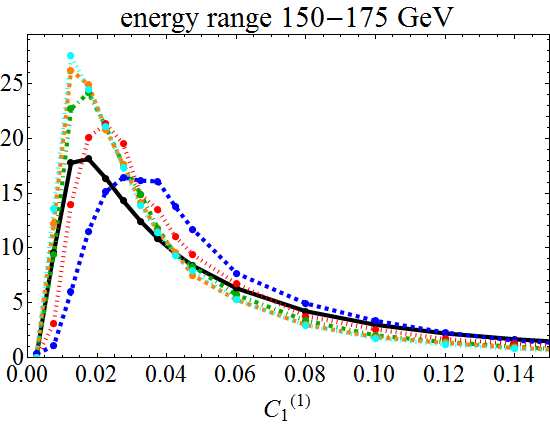}
\par\end{centering}
\vspace{-0.4cm}
\caption{Distribution of the jet opening angle $C_{1}^{(1)}$ for jets with energies in three
bins in pp collisions (black curves)~\cite{Larkoski:2014wba}; colored curves
show these distributions after an ensemble of jets has propagated
through the droplet of plasma, for different choices
of model parameters $a$ and $b$.
At small angles, each colored curve has been pushed to the right, to larger angles.  
At large enough angles, each colored curve has
been pushed down, which is equivalent to being pushed to the left.  
(For the blue curve this happens at larger angles than we have plotted.)
\label{fig:angular-distributions}}
\vspace{-0.2cm}
\end{figure*}

In order to have a distribution that mimics that jets  in pp collisions,
we choose our distribution of $E_{\rm jet}^{\rm init}$ and $\sigma_0$ such that the
distribution of jet energies is proportional to $(E_{\rm jet}^{\rm init})^{-6}$.
For our distribution of jet opening angles
for jets with a given $E_{\rm jet}^{\rm init}$,
we use 
the perturbative QCD calculations of variables denoted $C_{1}^{(\alpha)}$ that characterize
the angular shape of vacuum jets,
defined via~\cite{Larkoski:2013eya,Larkoski:2014wba} (see also Refs.~\cite{Banfi:2004yd,Jankowiak:2011qa})
\begin{equation}
C_{1}^{(\alpha)}\equiv\sum_{i,j}z_{i}z_{j}\left(\frac{|\theta_{ij}|}{R}\right)^{\alpha},\label{eq:C11}
\end{equation}
where $z_{i}$ is the fraction of the jet energy carried by hadron
$i$, $\theta_{ij}$ is the angular separation between hadrons $i$
and $j$, and $R$ is the radius parameter in the anti-$k_T$ reconstruction 
algorithm~\cite{Cacciari:2008gp}
used to find and hence define the jets.
Setting $\alpha=1$, the variable $C_1^{(1)}$ is a measure of the
opening angle of a jet.  We have no analogue of $R$ in our calculation,
since we have 
two known jets per ``event'', and hence 
no analogue
of jet finding or jet reconstruction.  Somewhat arbitrarily, we shall use $R=0.3$ in the definition
of $C_1^{(1)}$, since the jets in LHC heavy ion collisions whose angular shapes 
were measured in Ref.~\cite{Chatrchyan:2013kwa} were reconstructed with $R=0.3$.
The probability distribution for $C_1^{(1)}$ of quark and gluon jets with a given energy
$E_{\rm jet}^{\rm init}$  is given by equation
(A.8) in Ref.~\cite{Larkoski:2014wba},
where
it has been shown that these
distributions compare well with results from PYTHIA, and hence
with the distributions for jets produced in pp collisions. We shall use the distributions
for quark jets with $R=0.3$  in pp collisions with $\sqrt{s}=2.76$~TeV; some sample
curves can be seen as the solid curves in Fig.~\ref{fig:angular-distributions}.

There is no
rigorous connection between $\sigma_0$ and $C_{1}^{(1)}$:
our jets are not made of particles, so we have no 
fragmentation function and no 
$z_i$'s as in the definition (\ref{eq:C11}).   However, $C_{1}^{(1)}$ is
a measure of the opening angle of a jet in QCD and from Ref.~\cite{Chesler:2015nqz} 
we know that up to few percent corrections $\sigma_0$ is proportional to
the opening angle of the jet, defined there as the half-width at half maximum of the energy flux
as a function of angle.   Even without the further challenge
of connecting to $C_1^{(1)}$, the authors of Ref.~\cite{Chesler:2015nqz} advocate that the proportionality
constant in this relation should be seen as a free parameter, reflecting differences between jets in 
a confining theory like QCD and 
${\cal N}=4$ SYM.
We take
\begin{equation}
C_{1}^{(1)}=a\,\sigma_{0}\,, \label{eq:C11sigma}
\end{equation}
with $a$ the first of two free parameters 
in the specification of our model.
(A crude calculation, turning the angular distribution of the energy flux
in ${\cal N}=4$ SYM jets~\cite{Hatta:2010dz,Chesler:2014jva,Chesler:2015nqz}
into a fictional smooth distribution of many particles all carrying the same
small fraction of the jet energy, ignoring the caveats just stated,  and applying the
definition (\ref{eq:C11}) gives $a\sim1.7$.)


Finally, we describe the bulk AdS geometry, wherein the string will
propagate. We take a metric of the form 
\begin{equation}
ds^{2}=2\,dt\,dr+r^{2}\left[-f\left(r,x_\mu\right)dt+d\vec{x}_{\perp}^{\,2}+dz^{2}\right],\label{eq:metric-shock}
\end{equation}
with $r$ the AdS coordinate, and $(t,\vec{x}_\perp,z)$ the
field theory coordinates, with $z$ the beam direction. We take $f(r,x_\mu)=1-\left(\pi T(x_\mu)\,r\right)^{-4}$,
with $T(x_\mu)$ the temperature. This model neglects
viscosity and transverse flow.
For the temperature profile $T(x_\mu)$, we assume boost invariant longitudinal expansion 
(a simplification that makes the whole calculation boost invariant, meaning that we need only analyze jets with zero rapidity)
and use a simplified blast-wave expression 
for the transverse expansion~\cite{Ficnar:2013qxa}
\begin{equation}
T(\uptau,\vec x_\perp)=b\left[\frac{dN_{{\rm ch}}}{dy} \frac{1}{N_{\rm part}}\frac{\rho_{\text{part}}(\vec{x}_{\perp}/r_{\text{bl}}(\uptau))}{ \uptau 
\,r_{\text{bl}}(\uptau)^{2}}\right]^{1/3},\label{eq:temperature}
\end{equation}
where $\uptau\equiv\sqrt{t^2-z^2}$ is the proper time,
$\rho_{\text{part}}(\vec{x}_{\perp})$ is the participant density
as given by an optical Glauber model, $N_{\text{part}}\simeq 383$
and
$dN_{{\rm ch}}/dy\simeq 1870$ \cite{Abbas:2013bpa} are
the number of participants and 
the particle multiplicity at mid-rapidity in 2.76 ATeV 0-5\% centrality PbPb collisions at the LHC and $r_{\text{bl}}(\uptau)\equiv \sqrt{1+(v_{T}\uptau/R)^{2}}$,
with $v_{T}=0.6$ and $R=6.7$~fm. 
We initialize our calculation at $\uptau=1$~fm$/c$, neglecting the initial dynamics 
via which the hydrodynamic fluid formed and hydrodynamized. 
The constant $b$ is a
measure of the multiplicity per entropy $S$ and, for 
$S/N_{{\rm ch}}\simeq 7.25$~\cite{Muller:2005en,Gubser:2008pc} 
and $S/(T^3V)\simeq 15$~\cite{Borsanyi:2013bia,Bazavov:2014pvz} 
as in QCD at $T\simeq 300$~MeV, is given by $b\simeq 0.78$. 
(Note that $b=0.659$ in Ref.~\cite{Ficnar:2013qxa}.)
We shall treat $b$ as the second free parameter in our model
because the number of degrees of freedom is greater in ${\cal N}=4$ SYM theory than in QCD and the couplings in the theories differ too.
We are propagating ${\cal N}=4$ SYM
jets through an ${\cal N}=4$ SYM plasma with temperature $T$ meaning that we must
use a $b$ that is smaller than the QCD value.


\begin{figure*}[t]
\begin{centering}
\includegraphics[width=6.4cm]{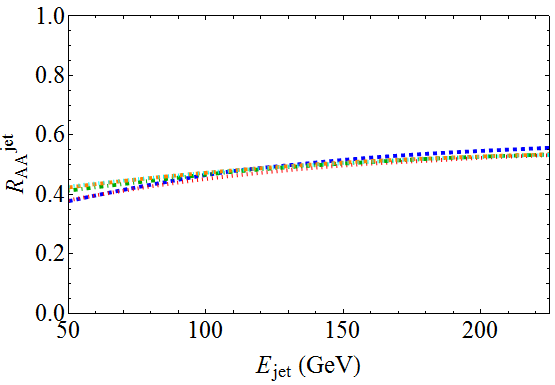}\quad\includegraphics[width=6.5cm]{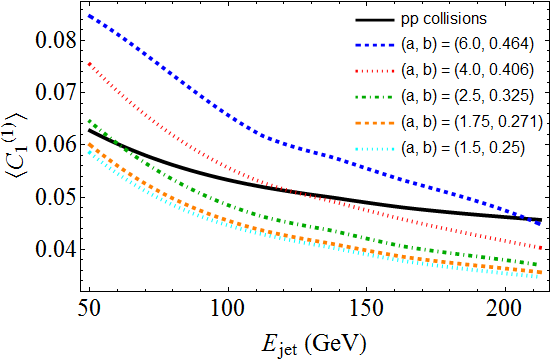}
\par\end{centering}
\vspace{-0.4cm}
\caption{Colored curves show $R_{AA}^{\rm jet}$ (left) and the ensemble average of the jet opening
angle, $\langle C_{1}^{(1)}\rangle$,
(right) for the final ensemble of jets after propagation 
through the droplet of plasma, for the
same combinations of $a$ and $b$ as in Fig.~\ref{fig:angular-distributions}. 
Here, $R_{AA}^{\rm jet}$ is the ratio of the number of jets
with a given energy after propagation through 
the plasma to that in the initial ensemble.  This quantity from our model should not
be compared quantitatively to experimental measurements of $R_{AA}$ for either hadrons or 
jets as we have no hadrons, no background, no
multi-jet events, and no jet finding or reconstruction. However,
we have chosen combinations of $a$ and $b$ such that
$R_{AA}^{\rm jet}$ is similar in all cases, and 
is similar to
$R_{AA}$ for jets in LHC heavy ion collisions~\cite{Raajet:HIN,Aad:2014bxa,Adam:2015ewa}. 
Even though $R_{AA}^{\rm jet}$ is so similar for all the colored curves,
the opening angle distributions (Fig.~\ref{fig:angular-distributions}) and their
mean $\langle C_{1}^{(1)}\rangle$ (right) vary significantly.  We have also plotted   $\langle C_{1}^{(1)}\rangle$ 
for the unperturbed ensemble,
as in pp collisions (black curve).  
\label{fig:raa}}
\vspace{-0.2cm}
\end{figure*}

As already noted, for simplicity
we initialize our jets at $\uptau=1$~fm$/c$, when we initialize
the plasma.  
%
%
We choose the initial position in the transverse
plane of our jets according to a binary scaling distribution,
proportional to $\rho_{\text{part}}(\vec{x}_{\perp})^{2}$, and choose
their transverse direction of propagation randomly.
For a given choice of our two model parameters $a$ and $b$,
we generate an ensemble of 
jets with their
initial position and direction distributed as just described and
their initial energy and opening angle 
($\sigma_0$ in the
dual gravitational description) 
distributed as described above.

We then allow each string in the ensemble to propagate in AdS,
as we have illustrated for a sample dijet in Fig.~\ref{fig:model}.
We compute the energy loss by integrating the string energy that falls
into the black hole (along the blue curves in Fig.~\ref{fig:model}) before
its temperature has fallen to a freeze-out temperature that we set
to 175~MeV (defined with $b=0.78$ so that our freeze-out time
is reasonable).  We assume that once the temperature has dropped
below freeze-out, the string that remains propagates in vacuum (along
the dashed grey curves in Fig.~\ref{fig:model}) meaning that the angle
at which the string endpoint travels downward into the AdS bulk no longer
changes.  
This final angle, which we denote $\sigma_*$, describes
the opening angle of the jet
that emerges from the droplet of plasma, $C_{1}^{(1)}=a\,\sigma_{*}$.
In this way, we extract the energy and opening angle of
each of the jets from among 
the initial ensemble that emerge from 
the droplet of plasma. We can then obtain the modified probability distribution 
of jet energies and opening angles in the final-state ensemble.
(We did the calculation by first evolving many tens of thousands of jets with varying values of 
their initial energy, opening angle, transverse position and direction and constructing
an interpolating function giving the final energy and final opening angle
as a function of these four input variables. We repeated this for each chosen value
of $b$.  For each value of $a$, we then reweighted these interpolating functions according to the
desired probability distribution for the four input variables, and sampled from the
reweighted final-state distributions in order to obtain the results that we shall present
below.)

Recapitulating, we have a two parameter model that describes how
the distribution of jet energies and opening angles changes due to
the propagation of the jets through an expanding, cooling droplet of
strongly coupled plasma relative
to what that distribution would be in a pp collision.
The model parameter $a$ sets the relationship between the downward angle followed by the string endpoint trajectory in the
gravitational description 
(initially, $\sigma_0$; after propagation through the plasma, $\sigma_*$)
and the jet opening angle $C_1^{(1)}$ --- whose initial distribution we have taken from perturbative QCD and
whose final distribution we have computed.
The model parameter $b$ controls the relationship between the temperature of the ${\cal N}=4$ SYM plasma in our model and
that of the QCD plasma we are modeling.  Larger $b$ results in a stronger gravitational pull on the string, meaning greater
energy loss and a greater increase in the opening angle of every jet in the ensemble.



\section{Results and Discussion}

We illustrate our results in Figs.~\ref{fig:angular-distributions} and \ref{fig:raa}
for five combinations of the model parameters $a$ and $b$. 
In Fig.~\ref{fig:angular-distributions}, we show how the probability distribution
for the jet opening angle $C_1^{(1)}$ is modified via propagation through
the plasma.  In Fig.~\ref{fig:raa}, we show that 
our
combinations of $a$ and $b$ 
each yield the same suppression
in the number of jets with a given energy in the final ensemble relative
to that in the initial ensemble, $R_{AA}^{\rm jet}$.   
The effect on 
$R_{AA}^{\rm jet}$
of increasing $a$ can be compensated by increasing $b$: 
increasing $a$ means reducing the $\sigma_0$ of the 
strings corresponding to jets with a given $C_1^{(1)}$; this reduces
their energy loss, which is compensated by increasing $b$. 
It is striking is how differently the 
$C_1^{(1)}$
distributions 
in  Fig.~\ref{fig:angular-distributions} and the
$\langle C_1^{(1)} \rangle$
in Fig.~\ref{fig:raa} 
are modified with the different combinations of $a$ and $b$.



There are two effects affecting
the probability distribution for the jet opening angle.   
First, as 
in
Fig.~\ref{fig:model}, each
null geodesic curves down, so all jets become wider~\cite{Chesler:2014jva,Chesler:2015nqz}. 
And, the larger $b$ is, meaning the larger the ${\cal N}=4$ SYM temperature $T$ in the calculation,
the stronger the gravitational force in AdS, the more the geodesics curve down, and the more
the jet opening angle distribution shifts to larger angle.   We see exactly this effect
in Fig.~\ref{fig:angular-distributions}, 
at all but large
values of the opening angle $C_1^{(1)}$.
Second, 
jets with a smaller $\sigma_0$ and hence
a smaller initial opening angle lose fractionally less energy~\cite{Chesler:2015nqz}.
This means that jets that initially had larger values of the opening angle $C_1^{(1)}$
lost more energy and got kicked out of the
energy bin corresponding to their panel in Fig.~\ref{fig:angular-distributions}, depleting this
large-angle region of the distribution.  This region of the distribution 
can get repopulated with jets that started out with substantially higher energy,
but because the initial energy distribution goes like $(E_{\rm jet}^{\rm init})^{-6}$ 
there are not enough of these jets to combat the depletion.  This depletion effect becomes
more significant the larger the value of $\sigma_0$, meaning that as the model parameter $a$ is 
reduced the $C_1^{(1)}$ above which the depletion
is significant comes down, as seen 
in Fig.~\ref{fig:angular-distributions}.
In Fig.~\ref{fig:raa} (right) we show the ensemble average
of the jet opening angle $C_1^{(1)}$.  We see that the combinations of $a$ and $b$ that
we have chosen that all yield comparable $R_{AA}^{\rm jet}$ can result in 
either one or the other of the two salient effects illustrated in Fig.~\ref{fig:angular-distributions}
being dominant, meaning that propagation through the plasma can result in
$\langle C_{1}^{(1)}\rangle$ increasing or decreasing.



There are of course many ways in which one could improve our model.
Collisions with nonzero impact parameter and nontrivial longitudinal
dynamics could be included, as could 
viscous hydrodynamics,
realistic transverse and longitudinal flow,
and
jets with nonzero rapidity. 
One could attempt to model effects on the jet of physics during the first fm$/c$ of the collision
and after freezeout, both of which we have neglected, or to consider an ensemble
of quark jets and gluon jets.
And, one can imagine
choosing probability distributions for other observables (dijet asymmetries; $C_1^{(\alpha)}$
for $\alpha \neq 1$) from data on pp collisions or perturbative QCD calculations and
studying how these distributions are modified in an ensemble of jets that
has propagated through the droplet of plasma produced in a heavy ion collision.

Our hope is that, even given its simplifications, our work can address
qualitative aspects of jet shape modifications, as for instance seen by CMS~\cite{Chatrchyan:2013kwa,CMS:2015kca}.
There, it is noticed that jets in heavy ion collisions are somewhat
narrower than jets with the same energy in pp collisions, if one focuses on particles
within the jets that are either close to the jet axis or have $p_T > 4$~GeV. 
Reconstructing jets 
incorporates soft particles at large angles originating from the wake 
of moving plasma trailing behind the jet rather than from the jet itself;
focusing on jet modifications at smaller angles
or higher $p_T$ therefore makes sense.
It is tempting to conclude that
the reduction in $\langle C_{1}^{(1)}\rangle$ due to the greater
energy loss suffered by jets with a larger initial opening angle may
be the dominant effect seen in these data.  This would point toward values of $a$ and $b$
in the lower half of the range that we have explored, where
the depletion at large angles dominates and the mean opening angle of jets with a given energy 
decreases even while every
jet in the ensemble broadens.


Remarkably, almost independent of the values of our model parameters our
model provides a clear qualitative prediction. When comparing the
angular distributions of pp collisions with AA collisions, as done
in Fig. \ref{fig:angular-distributions}, we see that the distribution
almost always has fewer jets with the smallest and the largest opening
angles, with the depletion at small angles due to the broadening
of the jets in the ensemble and the depletion at large angles originating
as described above.   Whether the mean opening angle goes up or 
down depends on which effect dominates but, regardless, we
expect the distribution of the opening angles of jets in AA collisions to be narrower
than in pp collisions.  
The striking qualitative features of the results we have already obtained from
our admittedly simplified model provide strong motivation for analyzing the distribution
of jet opening angles, as well as its mean, in other models for jet quenching, in Monte Carlo
calculations of jet quenching at weak coupling, and in analyses of data.




\noindent
{{\bf Acknowledgments:}}
We thank Jorge Casalderrey-Solana, Paul Chesler, Andrej Ficnar, Doga Gulhan, Simone Marzani, Guilherme Milhano, Daniel Pablos and Jesse Thaler
for useful discussions. We are especially grateful to Simone Marzani
for providing formulae from Ref.~\cite{Larkoski:2014wba}.
Research supported by the U.S. Department of Energy under grant Contract
Number DE-SC0011090.

\bibliographystyle{apsrev}
\bibliography{references}

\end{document}